\documentclass[aps,prl,twocolumn,groupedaddress,showpacs]{revtex4}
\usepackage{color}
\usepackage{amsfonts}
\usepackage{nccbbb}
\usepackage{graphicx}
\begin{document}

\title{Generation of pure bulk valley current in graphene}



\author{Yongjin Jiang$^{1,2}$}\email[]{jyj@zjnu.cn}
\author{Tony Low$^{3}$}
\author{Kai Chang$^{4,2}$}\email[]{kchang@semi.ac.cn}
\author{Mikhail I. Katsnelson$^5$}
\author{Francisco Guinea$^6$}
\affiliation{$^1$Center for Statistical and Theoretical Condensed Matter Physics and Department of Physics, ZheJiang Normal University, Zhejiang 321004, People's Republic of China  \\
$^2$Beijing Computational Science Research Center, Beijing 100084, China\\
$^3$ IBM T.J. Watson Research Center, Yorktown Heights, NY 10598, USA\\
$^4$ SKLSM, Institute of Semiconductors, Chinese Academy of Sciences, P.O. Box 912, Beijing 100083, People�s Republic of China \\
$^5$ Radboud University Nijmegen, Institute for Molecules and Materials,
Heyendaalseweg 135, 6525AJ Nijmegen, The Netherlands\\
$^6$ Instituto de Ciencia de Materiales de Madrid. CSIC. Sor Juana In\'es de la Cruz 3. 28049 Madrid, Spain
}



\date{\today}

\begin{abstract}
\color{black}{
The generation of valley current is a fundamental goal in 
graphene valleytronics but no practical ways of its realization are known yet.
We propose a workable scheme for the generation of bulk
valley current in a graphene mechanical resonator
through adiabatic cyclic deformations of the strains
and chemical potential in the suspended region.
The accompanied strain gauge fields can break the spatial mirror symmetry of the problem
within each of the two inequivalent valleys,
leading to a finite  valley current due to quantum pumping. 
An all-electrical measurement configuration is designed to detect the
novel  state with pure bulk valley currents.
}
\end{abstract}


\maketitle

Apart from pseudospin (chirality), charge carriers in graphene are also characterized by the valley index (sometimes called isospin)
originated from the existence of two conical (Dirac) points per Brillouin zone\cite{NetoRMP}. 
A valley polarized state requires the absence of time reversal symmetry, as the two valleys are related by this symmetry.

Motivated by the growing field of spintronics\cite{spintronics,valleytronics1}, it was proposed that the manipulation with the valley index may open a new way to transmit information through graphene, and different manipulation schemes were proposed\cite{Beenakker,DiXiao,topological confinement,trigonal,ZhenhuaWu-strain,GCN08,S10,Tony1,Golub}. After initial enthusiastic attitude, interest in ``valleytronics'' declined somehow, as it was soon realized that a valley polarized current will be degraded by intervalley scattering induced by atomic scale disorder\cite{NetoRMP}, making it difficult the maintenance of valley polarized states. In addition, a number of proposals were based on the spatial separation of valley currents at zigzag edges\cite{Beenakker}, which requires well-defined edge orientation and at the same time free of short-range scattering.

We present a new scheme to induce valley polarized currents in graphene which avoids some of the pitfalls of previous proposals. The breaking of time reversal symmetry is achieved by means of time dependent fields, instead of a magnetic field. The induction of valley polarization by $a.c.$ fields has been proven in MoS$_2$\cite{XLFXY12} as well as in (111)-oriented silicon metal-oxide-semiconductor
field-effect transistors\cite{Karch} where optical radiation was used in order to excite valley polarized charge carriers. As in these experiment, the scheme discussed below generates valley currents throughout the entire system. However, the $a.c.$ driving force in our proposal is due to mechanical vibrations of a nanoelectrical-mechanical system (NEMS\cite{resonator1,resonator2,resonator3,resonator4}), as
illustrated in Fig.1a.


\begin{figure}[t]
\scalebox{0.5}[0.5]{\includegraphics*[viewport=170 180 640 350]{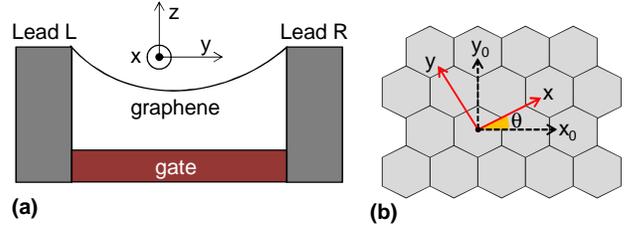}}
\caption{Illustration and definitions
of the \textbf{(a)} graphene based NEM
device and \textbf{(b)} crystallographic
coordinate systems used in the paper.}
\label{fig1}
\end{figure}

We employ a continuum-medium description of
graphene with the Dirac Hamiltonian,
\begin{equation}
{\cal H}^{\pm}(\vec{k},\vec{{\cal A}})=\hbar v_f[\pm (k_x\mp {\cal
A}_x)\sigma_x+(k_y\mp {\cal A}_y)\sigma_y]+\mu \bbbone_2,
\label{eq:1}
\end{equation}
where $+(-)$ denotes $K$($K'$) valley index, $v_f$ is the Fermi
velocity and $\mu$ is the chemical potential. $\vec{{\cal
A}}(\vec{r})$ is an effective gauge field describing the
modifications to the hopping amplitudes induced by the strain
fields $u_{ij}(\vec{r})$ \cite{Ando2002,gaugePhysRep}, and has opposite signs at the two
valleys as required by time-reversal symmetry. We described the
deformation of the suspended region, i.e.
$-\frac{L}{2}<y<\frac{L}{2}$, with a simple uniaxial strain given by
$u_{yy}=u$ and $u_{xx}=u_{xy}=0$. It will be shown below that for
arbitrary crystallographic orientation $\theta$ (see Fig.1b), $\vec{{\cal
A}}(\vec{r})$ in the suspended region is given by the expression
\begin{equation}
\vec{{\cal A}}(\vec{r})=\frac{\beta \kappa
u}{a}(\mbox{-cos}3\theta,\mbox{sin}3\theta),
\label{eq:A-strain}
\end{equation}
where $\beta\approx 2$ is the electron Gruneisen parameter  and $\kappa\approx \frac{1}{3}$ is a parameter related to graphene's elastic property as described in \cite{Ando2002}.
$a\approx 1.4\AA$ is the interatomic distance. Adiabatic cyclic
variations of the internal parameters, such as deformations in the
strains ($u$) and chemical potential ($\mu$) in the suspended
region, over a work cycle can constitute a scheme for adiabatic quantum
pumping\cite{PSS09,pumping1} (for the general theory of quantum pumping, see Ref. \cite{Brouwer98,Buttiker2002}).
For the charge pumping one needs to break the spatial mirror symmetry, e.g., making
the right and left leads different (e.g., by different doping)\cite{Buttiker2002}.
Here we will demonstrate that for the case of {\it symmetric} leads the {\it valley}
current through the system is, in general, nonzero. In this case, 
the pumping current through each channel will be
shown to follow the general relation (with periodic boundary condition 
along the transversal $x$ direction),
\begin{equation}
\nonumber
I_{L/R}^{pump,K}(k_x,\theta)=-I_{L/R}^{pump,K'}(-k_x,\theta)\neq 0 
\mbox{\,\,\,\,\,if\,\,\,\,\,} \theta\neq n\frac{\pi}{3}\\
\label{central-result}
\end{equation}
where the subscript $L/R$ refer to left/right leads ($I_{L/R}$ is defined postive when the current flows out of the device). 
Eq.(\ref{central-result}) represents the central result of our paper, and
embodies the intrinsic symmety of valley pumping in graphene NEMs, as we will show later. 
Below, we present detail derivations leading to
Eq.(\ref{central-result}) and discuss the physical consequences that
follows, such as the generation of pure bulk valley current and its
possible experimental detection.a

We start by considering the case $\theta=0$, where the trench
is directed along the zigzag direction (see the $(x_0,y_0)$ coordinate system shown in Fig.\ref{fig1}b).
In this case,  the Hamiltonian
has the form of Eq.(\ref{eq:1}) and the pseudo-magnetic vector potential reads:
\begin{eqnarray}
{\cal A}_x&=&\frac{\beta \kappa}{a}(u_{xx}-u_{yy}),\nonumber\\
{\cal A}_y&=&-\frac{2\beta \kappa}{a}u_{xy}. \label{eq:2}
\end{eqnarray}
For arbitrary orientation i.e.  $\theta\ne 0$, Eq.(\ref{eq:1}) and Eq.(\ref{eq:2}) have to
be recasted in the new coordinate frame $(x,y)$. The two coordinate system are related by
\begin{eqnarray}
\left(%
\begin{array}{c}
    x \\
    y \\
\end{array}%
\right)={\cal R}\left(%
\begin{array}{c}
    x_0\\
    y_0 \\
\end{array}%
\right) \,\,\,\,\,\,\,\mbox{,}\,\,\,\,\,\,\,
{\cal R}=\left(%
\begin{array}{cc}
    \cos\theta&\sin\theta \\
    -\sin\theta &\cos\theta\\
\end{array}%
\right).
\end{eqnarray}
The wave vector $\vec{k}$ transforms in the
same way as $\vec{r}$ such that $\vec{k}\cdot \vec{r}$ is a rotational
invariant quantity. The strain field is defined as,
$u_{ij}=\frac{1}{2}(\frac{\partial u_j}{\partial
x_i}+\frac{\partial u_i}{\partial x_j})$, which is a symmetric
tensor of rank two. Hereafter we use the subscript/superscript ``0"
to denote physical quantities in the original frame $(x_0,y_0)$.
Thus, we have $\vec{k}_{0}=R^{-1}\vec{k}$
and
\begin{eqnarray}
u_{xx}^0-u_{yy}^0&=&\cos2\theta(u_{xx}-u_{yy})-2\sin2\theta u_{xy}, \nonumber\\
u_{xy}^0&=&\frac{1}{2}\sin2\theta(u_{xx}-u_{yy})+\cos2\theta
u_{xy}.\label{eq:straintransform}
\end{eqnarray}
By using the new coordinates in the Dirac Hamiltonian, we can transform it to the rotated frame:
\begin{widetext}
\begin{eqnarray}
 {\cal H}^{\pm}(\vec{k},\vec{{\cal A}})=\hbar v_f\left(\begin{array}{cc}
 \mu/\hbar v_f & e^{\mp i\theta}[(\pm k_x-ik_y)\mp(\pm {\cal A}_x-i{\cal A}_y)]  \\
e^{\pm i\theta}[(\pm k_x+ik_y)\mp(\pm {\cal A}_x+i{\cal A}_y)]) & \mu/\hbar v_f\\
\end{array}\right),  \label{eq:15}
\end{eqnarray}
\end{widetext}
where we have defined the pseudo-magnetic field $\vec{\cal A}$ in the
rotated frame as\cite{Fzhai}:
\begin{eqnarray}
\left(\begin{array}{c}
  {\cal A}_x \\
{\cal A}_y  \\
\end{array}\right)={\cal R}(3\theta)\left(\begin{array}{c}
  \frac{\beta \kappa}{a}(u_{xx}-u_{yy}) \\
\frac{-2\beta \kappa}{a}u_{xy} \\
\end{array}\right). \label{eq:define A}
\end{eqnarray}
Eqs.(\ref{eq:15}) and (\ref{eq:define A}) constitute the
continuum-medium description of stained-graphene in an arbitrarily rotated frame.
Here follows two comments about this general form. First, if we
perform the gauge transformation of the wave function on $B$ sublattice $\psi_B^{\pm K}\longrightarrow
\psi_B^{\pm K}e^{\pm i\theta}$, the Dirac Hamiltonian can be made
explicitly invariant (i.e. Eq.(\ref{eq:1})) under rotation. Thus, we can simply drop the factor
$e^{\pm i\theta}$ in Eq.(\ref{eq:15}) in subsequent discussion.
Second, from the definition for pseudo-magnetic field, it is
obvious that the form is of $2\pi/3$ periodic in $\theta$, which reflects
the trigonal symmetry of the underlying honeycomb lattice.

Next, we describe the quantum pumping problem based on
graphene NEMs\cite{Fogler,pumping1}. As discussed above, the derived Hamiltonian given by Eq.(\ref{eq:15}) is physically equivalent
to Eq.(\ref{eq:1}). Suppose that between
$-\frac{L}{2}<y<\frac{L}{2}$, the system has uniaxial strain
$u_{yy}=u$ and $u_{xx}=u_{xy}=0$, hence,  Eq.(\ref{eq:define A}) is reduced to
\begin{equation}
({\cal A}_x,{\cal A}_y)(\vec{r})=\left\{%
\begin{array}{ll}
     0, & \hbox{$|y|>\frac{L}{2}$} \\
   (\frac{\beta \kappa u}{a})(-\cos3\theta,\sin3\theta), & \hbox{$|y|\leq \frac{L}{2}$} \\
\end{array}%
\right.
\label{vector potential}
\end{equation}
with the expression given in Eq.(\ref{eq:A-strain}) for the suspended region.
In the following, assuming a particular geometry (the width $W$$\gg$$L$), the $x$ direction is treated as translationally invariant.
From Eq.(\ref{eq:15}) and Eq.(\ref{vector potential}), one can easily see that
\begin{eqnarray}
{\cal H}^{+}(k_x,k_y,\vec{{\cal A}},\theta)={\cal H}^{-}(-k_x,k_y,\vec{{\cal A}},-\theta).
\label{symmetry}
\end{eqnarray}
We can call such combined symmetry as the
\emph{crystalline-angle-combined mirror symmetry} in the continuum-medium
model. 
It turns out that such combined symmetry has a
significant consequence on the relation of pumping currents in each valley, i.e.,  Eq.(\ref{central-result}), as will be
elaborated further.

Due to the mentioned symmetry, we may focus only on the $K$ Dirac cone, whose Hamiltonian has the following
form for the different regions (i.e., $i=L,R,G$ denotes
$y<-\frac{L}{2}$, $y>\frac{L}{2}$, $|y|\leq\frac{L}{2}$,
respectively):
\begin{eqnarray}
{\cal H}_i^{+}(\vec{k},\vec{{\cal A}})&=&\hbar v_f\left(\begin{array}{cc}
 \epsilon_i/\hbar v_f&(\tilde{k}_x-i\tilde{k}_y)_i  \\
(\tilde{k}_x+i\tilde{k}_y)_i & \epsilon_i/\hbar v_f \\
\end{array}\right),\label{eq:21}
\end{eqnarray}
where we have
defined $(\tilde{k}_x,\tilde{k}_y)_i=(k_x-{\cal A}_x,k_y-{\cal A}_y)_i$.
In this paper, we consider mainly the symmetric case $\epsilon_L=\epsilon_R\neq\epsilon_G$.

The eigenenergies and eigenstates of the Hamiltonian in Eq.(\ref{eq:21}) read
 \begin{eqnarray}
 E_{i}(\vec{k})&=&\epsilon_i+s\hbar v_f\sqrt{\tilde{k}_x^2+\tilde{k}_y^2},\nonumber\\
 \psi_{i}(\vec{k})&=&e^{i(k_xx+k_{y_i}y)}\left(\begin{array}{c}
 1 \\
 \frac{\hbar v_f(\tilde{k}_x+i\tilde{k}_y)_i}{E_s-\epsilon_i} \\
\end{array}\right),\label{eq:eigen}
 \end{eqnarray}
where $s=\pm 1$ refers to
electron/hole band, respectively. Due to translational invariance in $x$ direction, $k_x$ is the same
in all three regions. We consider now the equilibrium situation where all three regions can be
described by a common chemical potential $\mu$. Obviously,  $\tilde{k}_{x_i}$ is always real since $k_x$ is
real. Then,
$\tilde{k}_{y_i}=\pm\sqrt{k_{f_i}^2-\tilde{k}_{x_i}^2}$, where
$k_{f_i}=\frac{\mu-\epsilon_i}{\hbar v_f}$ and the $\pm$ sign  is selected to give the correct sign in the group velocity,
depending whether it is an incident, transmitted, or reflected waves.
Note that $\tilde{k}_y$ can be purely imaginary representing evanescent
waves in the central region.

It is straightforward to calculate the scattering matrix for our device. Without loss of generality, we can focus on the case with
the electron doping ($\mu-\epsilon_L>0$) in the leads. The scattering
coefficients are calculated to be
\begin{eqnarray}
r_d&=&e^{-ik_{y_L}L}\frac{C_2(d)+C_3(d)}{C_1(d)},\nonumber\\
t_d&=&e^{-ik_{y_L}L}\frac{-4\sin\varphi_L\sin\varphi_Ge^{iA_yLd}}{C_1(d)},\label{eq:scatteringmatrix}
\end{eqnarray}
where $r_d$ and $t_d$ are the reflection and transmition
coefficients with $d=1(-1)$ corresponding to the cases with
incident waves from $y=-\infty$ (left lead) and $y=\infty$ (right lead), respectively;
$\varphi_L$ and $\varphi_G$ are defined through
\begin{eqnarray}
e^{\pm i\varphi_L}=\frac{k_x\pm ik_{y_L}d}{k_{f_L}},e^{\pm i\varphi_G}=\frac{\tilde{k}_{x_G}\pm i\tilde{k}_{y_G}}{k_{f_G}}.
\end{eqnarray}
The $C_i(d)$ in Eq.(\ref{eq:scatteringmatrix}) are defined
as 
\begin{eqnarray}
C_1(d)&=&4i\sin(\tilde{k}_{y_G}Ld)(1-\cos\varphi_L\cos\varphi_G)\\
&-&4\cos(\tilde{k}_{y_G}Ld)\sin\varphi_L\sin\varphi_G,\nonumber\\
C_2(d)&=&-2i(1+e^{2i\varphi_L})\sin(\tilde{k}_{y_G}Ld),\nonumber\\
C_3(d)&=&2ie^{i\varphi_L}[\sin(\tilde{k}_{y_G}Ld+\varphi_G)+\sin(\tilde{k}_{y_G}Ld-\varphi_G)].\nonumber
\label{eq:C functions}
\end{eqnarray}

\emph{Symmetries related with $t_d$ and $r_d$}. Next, we
discuss symmetry properties of the scattering amplitudes. First, we note that
$\varphi_G(d)=\varphi_G(-d)$, $\varphi_L(d)=-\varphi_L(-d)$ and
they are independent of the sign of
$\theta$. Then, we can obtain the relations satisfied by $C_i$'s: $C_1(d)=-C_1(-d)$, $C_{2,3}(-d)^*=C_{2,3}(d)$.
Based on these relations and the odd parity of  ${\cal A}_y(\theta)$, we arrive at:
\begin{eqnarray}
t_d(k_x,\theta)&=&t_{-d}(k_x,-\theta),\nonumber\\
r_d(k_x,\theta)&=&r_d(k_x,-\theta). \label{eq:t-finite-theta}
\end{eqnarray}

\emph{Symmetry of pumped valley-dependent current}.  According to the adiabatic pumping theory\cite{Brouwer98}
and the symmetry relations satisfied by $r_d$
and $t_d$,   we obtain the following relation for the pumping current $I^{pump,K}_L(k_x,\theta)$ for $K$ valley:
  \begin{eqnarray}
&&I^{pump,K}_L(k_x,\theta)-I^{pump,K}_R(k_x,-\theta)\nonumber\\
&=&\frac{ie\omega}{4\pi^2}\int_{0}^{2\pi/\omega}ds[\frac{dr_1(\theta)}{ds}r_1(\theta)^*-\frac{dr_{-1}(\theta)}{ds}r_{-1}(\theta)^*],\,\,\,\,\,\,\,\,\,\,\,\,\,\,,\label{eq:pump1}
\end{eqnarray}
where we've used $s$ as time symbol to avoid confusion with transmission coefficient.  Now, using the symmetry relations satisfied
by $C_i$'s and the fact that the common complex factors for
$C_2$ and $C_3$, i.e., $ie^{i\varphi_L}$, is independent of time(see Eq.(\ref{eq:time depencence}) for typical time dependence of pumping parameters for graphene NEM), 
we can simply prove that the integrand in Eq.(\ref{eq:pump1}) is zero. By further taking into account the current conservation
condition( $I^{pump,K}_L(k_x,-\theta)+I^{pump,K}_R(k_x,-\theta)=0$)
and the symmetry relation guaranteed by Eq.(\ref{symmetry}) (i.e., $I^{pump,K'}_L(-k_x,\theta)=I^{pump,K}_L(k_x,-\theta)$), we arrive at the first part of
Eq.(\ref{central-result}), the relation for pumped valley-dependent current, which is the central result of this paper.
The above derivation based on the explicit solution of scattering amplitudes demonstrates the usefulness of the generic crystalline-angle-combined mirror symmetry of the suspended graphene under uniaxial strain. Later we will show that Eq.(3) and its consequences are in consistent with inversion symmetry of the whole system.

\begin{figure}[tbp]
\scalebox{0.3}[0.3]{\includegraphics*[viewport=0 10 770 590]{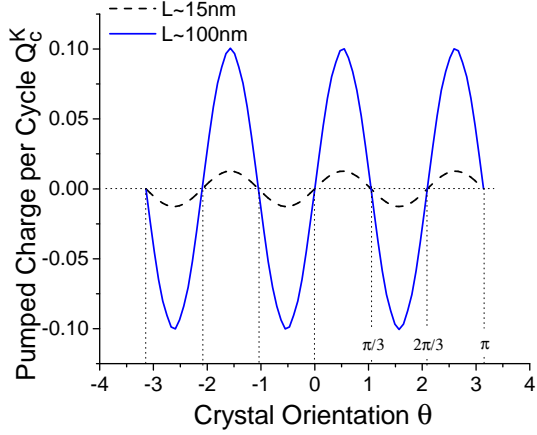}}
\caption{(Color online)
The pumped charge per cycle of the $K$ valley,
denoted as $Q^{K}_c$. The valley current can be obtained by
$e\omega Q^{K}_c/2\pi$. The width is fixed to be 5000$a$ ($\approx
700nm$), calculated for two different length as indicated.
We fixed the phase difference of the driving parameters $\phi=\pi/2$
and take $|\vec{A}_{av}|/k_f\approx 0.002$, which amounts
to a strain $u\approx 1.75\times 10^{-4}$.}
\label{fig2}
\end{figure}

Next, we discuss some direct consequences of Eq.(\ref{central-result}).
By integrating out the transversal wave vector $k_x$, we can get the following relation,
\begin{eqnarray}
I^{pump,K}_{L/R}(\theta)=-I^{pump,K'}_{L/R}(\theta)\label{eq:pumpI-antisym}.
\end{eqnarray}
Eq.(\ref{eq:pumpI-antisym}) means that the total pumping current at a given lead is opposite for different valleys. Thus,
the total charge current is exactly zero. 
This situation is analogous to the pure spin current generation in spintronics\cite{spincurrent1,spincurrent2},
thus we call this effect pure valley current generation. In summary, we have shown that the application of an alternating back gate voltage to graphene NEMs can induce a pure valley current via adiabatic pumping. In adiabatic pumping theory, there are two necessary conditions for finite charge pumping effect, i.e.,  time reversal symmetry breaking and mirror symmetry breaking of the the left/right leads\cite{Buttiker2002}. The above derivations show that the valley pumping effect can be
realized in a \emph{seemingly} symmetric two-terminal
geometry with identical leads. However, the mirror symmetry of the system
(i.e. $y$$\rightarrow$$ -y$) has actually been broken for each valley when $\theta\ne \frac{n\pi}{3}$.
Such symmetry breaking is embodied in the continuum theory through a nonzero
${\cal A}_y$ component. This constitutes the second part of Eq.(\ref{central-result}).
For more quantitative understanding, we present some numerical results of $I^{pump,K}_L(\theta)$
(in terms of the pumped charge per cycle ) for the $K$ valley in Fig.\ref{fig2}
using some typical experimental parameters, as discussed below.

As stated above,
the strain ($u$) and Dirac potential ($\epsilon_G$) in the suspended
region are modulated by the ac back gate voltage.
Near resonance, they differs by a phase difference  $\phi$ with a typical time dependence given by(with conventional time symbol $t$)\cite{pumping1}:
\begin{eqnarray}
\epsilon_G&=&\epsilon_{G0}[1+\delta\epsilon\cos(\omega t)]^{\frac{1}{2}},\nonumber\\
\vec{{\cal A}}&=&\vec{{\cal A}}_{0}[1+\delta {\cal A}\cos(\omega t+\phi)]^2.
\label{eq:time depencence}
\end{eqnarray}
Assuming typical numbers for the static part $\epsilon_{G0}=\epsilon_{L}=0.3eV$,
$\vec{{\cal A}}_{0}=0.02k_f(\cos3\theta,-\sin3\theta)$, and the
oscillating amplitude $\delta\epsilon=\delta {\cal A}=0.2$ , we calculated the pumped charge per cycle $Q^K_c(\theta)$  for
$K$ valley, as shown in Fig.2. By definition, the pumping current $I^{pump,K}_L(\theta)=e\omega Q^K_c(\theta)/2\pi$.
Our calculation indicates a nearly linear scaling(unshown) of the pumping effect on length $L$ of the NEMs, which is similar to results in\cite{pumping1}. As explicitly shown,
the maximum pumping effect is reached for the crystallographic angles corresponding to
armchair-type $x$ axis ($\theta=\pi/2+n\pi/3,n\in Z$) while it is zero at zigzag-type
$x$ axis ($\theta=n\pi/3,n\in Z$). The periodicity $2\pi/3$ with $\theta$ is easily seen.
The valley current can
be defined as $I^{pump,v}(\theta)=I^{pump,K}(\theta)-I^{pump,K'}(\theta)$, which
is simply twice the value of $I^{pump,K}(\theta)$. For typical resonance frequency of $\omega\approx$ 10$MHz$\cite{resonator3} to 0.16$GHz$
\cite{resonator4},
we arrive at numerical estimates $I^{pump,v}\approx 0.1-10\,$pA/$\mu$m, a quantity measurable in experiment. Furthermore, the signal can be amplified by (i) increasing the area of suspended graphene part (the signal is roughly proportional to both length and width in the theoretical model we considered)  and (ii) tuning the amplitude of the ac voltage of the back gate to increase the oscillation  magnitude of strain/chemical potential. 

\begin{figure}[t]
\scalebox{0.4}[0.4]{\includegraphics*[viewport=120 190 680 450]{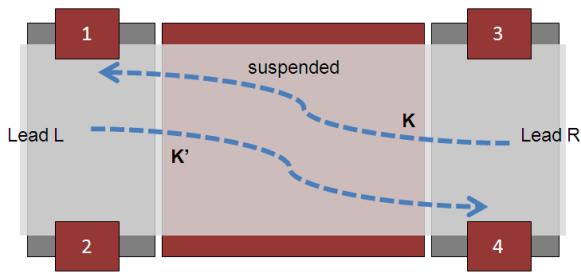}}
\caption{(Color online) Pictorial sketch of the pumping generation of pure valley current and an all-electrical detection scheme.
The Hall voltage difference $V_{12}=V_{1}-V_{2}$ and
$V_{34}=V_{3}-V_{4}$ across the NEMs is predicted to bear opposite
sign due to the flow of pure valley
current.}\label{figure:electrical measurement}
\end{figure}

The possibility of pure bulk valley current in this simple pumping
scenario looks very promising. The problem is how to probe the valley current. 
Here we propose an all-electrical
measurement as shown in Fig.\ref{figure:electrical measurement}(note the voltage contacts are patterned on the supporting leads, instead of on the suspended region).
From Eq.(\ref{central-result}), we can infer that the charge current
pertaining to carriers from the valley $K$  not only has opposite
longitudinal component with respect to the charge carriers from the valley $K'$,
but also their transversal velocities are opposite.  As
pictorially shown in Fig.\ref{figure:electrical measurement}, we
expect charges accumulating on opposite edges on the two sides of
the NEMs. Based on this observation, we predict that the resultant Hall
voltage on the left lead has opposite sign with that on the right
lead, i.e., $\mbox{Sign}(V_{12}/V_{34})=-1$, which is the characteristic
feature of the bulk valley current flow.

The above picture can be made more quantitative. Because $k_x$, like charge, is a conserved quantum number
 in our pumping scenario,  we can introduce  a quantity, \emph{valley-dependent pumping  Hall current},
which can be calculated as $I^{Hall,K}_{L/R}(\theta)=\sum_{k_x}I^{pump,K}_{L/R}(k_x,\theta)\frac{k_x}{k_y}$. This  Hall current accompanying the pure bulk valley current can be viewed as resulted from an effective pumping force, instead of the usual Lorentz force due to a magnetic field.  From Eq.(\ref{central-result}) and the particle conservation law for each $k_x$ channel, we can obtain:
\begin{equation}
I^{Hall,K}_{L/R}(\theta)=I^{Hall,K'}_{L/R}(\theta)=-I^{Hall,K}_{R/L}(\theta).
\end{equation}
Such Hall current on different leads can result in the opposite Hall voltage. 
Thus the existence of pure valley currents possibly can be detected in the leads by means of nonlocal multiterminal measurements\cite{Aetal11}. Finally, it's worthy to point out that
 such hall current pattern is a reasonable consequence of inversion symmetry and time reversal symmetry breaking of the system.


Our discussion above is restricted to the case with symmetrical leads. It is straightforward to extend our study to the more general  case with differently doped leads.
In the general situation, the current is not purely valley current, i.e., $I^{pump,K}_{L/R}(\theta)\neq -I^{pump,K'}_{L/R}(\theta)$, thus the charge pumping current is finite.

To conclude, we have shown that through pumping induced by mechanical vibrations
bulk valley polarized currents can be generated in graphene leads connecting
the graphene resonator with trench directed at a general
crystallographic angle. We have demonstrated that the generated 
current is purely valley current (with zero net charge pumping current) in the setup
with the same doping rate in the graphene leads. Together with an
all-electrical measurement scheme, our proposal opens a new
direction of exploiting the valley degree of freedom, thus pushing
forward graphene-based valleytronics a step forward toward real
applications.

Y. J. and K. C. acknowledge the support from the
National Natural Science Foundation of China [under
Grants No. 11004174 (Y. J.) and No. 10934007 (K. C.)].
Y. J. is also supported by the Program for Innovative
Research Team in Zhejiang Normal University. K. C. is
also supported by the National Basic Research Program of
China (973 Program) under Grant No. 2011CB922204.
T. L. also acknowledges partial support from NRIINDEX.
M. I. K. acknowledges financial support from
FOM, Netherlands. F. G. acknowledges financial support
from Spanish MICINN (Grants No. FIS2008-00124,
No. FIS2011-23713, and No. CONSOLIDER CSD2007-
00010), and ERC, Grant No. 290846.

\end{document}